\documentclass[twocolumn,twoside,slac_two]{revtex4}
\usepackage{graphicx}
\usepackage{fancyhdr}
\renewcommand{\headrulewidth}{0pt}
\renewcommand{\footrulewidth}{0pt}

\begin{document}

\title{Future e$^{+}$e$^{-}$ Flavour Factories:
accelerator challenges}

\author{M. E. Biagini}
\affiliation{INFN, Laboratori Nazionali di Frascati, Frascati (RM),
Italy}

\begin{abstract}

Operation of the B-Factories (PEP-II and KEKB) has been very
successful, both having exceeded their design peak and integrated
luminosity and provided a huge amount of good data to the
experiments. Proposal for upgrades, in order to achieve about two
order of magnitude larger luminosity, are in progress in Japan, with
Super-KEKB, and in Europe, with SuperB. Very high beam intensity,
very short bunch length and low Interaction Point $\beta$-functions
are the key points of the Japanese design, very challenging for the
hardware components (RF, vacuum). On the other hand SuperB exploits
a new collision scheme, namely large Piwinski angle and ``crab
waist", which will allow for reaching a luminosity two order of
magnitude larger without increasing beam currents and decreasing
bunch lengths. In this talk the present status of the two projects
will be briefly reviewed.
\end{abstract}

\maketitle

\thispagestyle{fancy}

\section{Introduction}

\begin{figure*}[t]
\centering
\includegraphics[width=100mm]{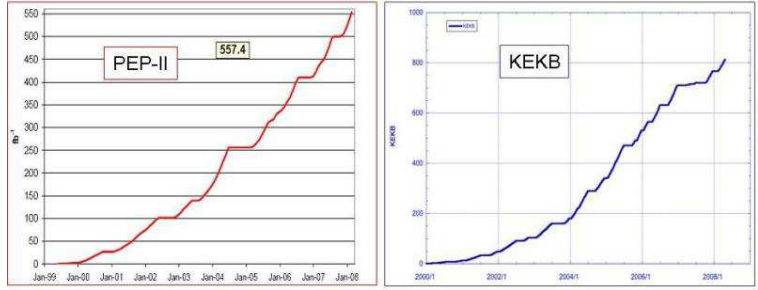}
\caption{Integrated luminosity for PEP-II (left) and KEKB (right) at
end of April 2008.} \label{Fig1}
\end{figure*}

Presently operating B-Factories (PEP-II and KEKB) have exceeded
their design goals, both in peak and integrated luminosity. PEP-II
\cite{PEPII}, running from mid-1999 to April 2008, has reached 4
times the design peak luminosity, delivering to the BaBar experiment
an integrated luminosity larger than 557 fb$^{-1}$ (see Fig. \ref{Fig1}, left plot). KEKB \cite{KEKB} also started operation in
1999 and reached a peak luminosity 60$\%$
 higher than the design value, delivering about 820 fb$^{-1}$
 (up to April 2008) to Belle (see Fig. \ref{Fig1}, right plot).
 In Table \ref{T1} the performances reached at the end of April 2008 are summarized.
Very good performances and high operation reliability represent a
big success for all the Factories, and upgrade of an order of
magnitude or more in luminosity is desirable for investigation on
particle physics beyond the Standard Model.

\begin{table}[h]
\begin{center}
\caption{B-Factories performances (April 2008).}
\begin{tabular}{|l|c|c|}
\hline \textbf{ } & \textbf{PEP-II} & \textbf{KEKB}\\
\hline Energy (GeV)
 & 3.1x9 & 3.5x8 \\
\hline Design peak L ($\times$10$^{33}$cm$^{-2}$s$^{-1}$)
 & 3 & 10 \\
\hline Achieved peak L ($\times$10$^{33}$cm$^{-2}$s$^{-1}$)& 12 & 17 \\
\hline Design int. L/day (pb$^{-1}$)  & 130 & 600 \\
\hline Achieved int. L/day (pb$^{-1}$) & 911 & 1231 \\
\hline Achieved total int. L (fb$^{-1}$) & 557.4 & 824 \\
\hline
\end{tabular}
\label{T1}
\end{center}
\end{table}

The construction and operation of multi-bunch e$^{+}$e$^{-}$
colliders have brought about many advances in accelerator physics
 in the area of high currents, complex interaction regions,
  high beam-beam tune shifts, high power RF systems, controlled beam
  instabilities, rapid injection rates, and reliable uptimes (about 95$\%$).
   The present B-Factories have proven that their design concepts are valid,
   since asymmetric energies work well, the beam-beam energy transparency
   conditions are weak, high currents can be stored and the electron cloud
   instability (ECI) can be managed.
On the detector-machine side the Interaction Regions (IR)
backgrounds can be handled successfully and IR with two energies can
work. Moreover unprecedented values of beam-beam parameters have
been reached (0.06 up to 0.09), and continuous injection in
production has helped increasing the integrated luminosity. However
a step forward is needed in order to increase luminosity by one or
even two order of magnitude.

\section{Two Approaches}

To increase Luminosity of about two orders of magnitude, with the
same philosophy of the present B-Factories, borderline parameters
are needed such as those chosen by the Super-KEKB project, that is:

\begin{itemize}
      \item very high currents;
      \item smaller $\beta_y^*$;
      \item smaller damping times;
      \item very short bunches;
      \item crab cavities for head-on collision;
      \item higher power.

\end{itemize}

To squeeze the vertical beam size, so increasing Luminosity, the
vertical $\beta_y^*$ at the Interaction Point (IP) must be
decreased: this is efficient only if at the same time the bunch
length is shortened to about the $\beta_y^*$ value, otherwise
particles in the head and tail of the bunch will see a larger
$\beta_y^*$ (hourglass effect). However shorter bunches require an
increase of RF voltage with consequent costs increase. This approach
is then difficult in terms of operational costs because of the large
RF power needed, the higher backgrounds, and High Order Modes (HOM)
heating.

The SuperB project exploits an alternative approach, with a new
collision scheme \cite{Raimondi}:

\begin{itemize}
      \item very small beams (ILC-Damping Rings like);
      \item large Piwinski angle and ``crab waist";
      \item currents comparable to present Factories.
\end{itemize}

This configuration moves the difficulties to the realization and
maintenance of extremely focused beams. Remarkably, SuperB would
produce this very large improvement in luminosity with circulating
currents and wall plug power similar to those of the current
B-Factories.

Both approaches require status-of-the-art technology, but the
operation of SuperB would probably be easier.

\section {KEKB and Super-KEKB}

\begin{figure}[h]
\centering
\includegraphics[width=60mm]{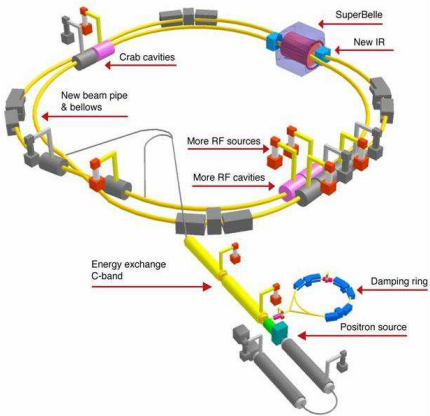}
 \caption{Layout of Super-KEKB.} \label{Fig2}
\end{figure}

Since 2004 a major upgrade of KEKB has been studied, and has been
described in a Letter of Intent \cite{SuperK}. A layout of the
machine is in Fig. \ref{Fig2}.

What are the challenges of the Super-KEKB design is straightforward
when looking at the simplified luminosity formula below:

\begin{equation}
  L \approx  { { \gamma _{\pm}} \over {2 e r_{e} }}\ {{I_{\pm}\xi_{\pm y}} \over {\beta^{*}_{y}}}
\label{eq:luminosity}
\end{equation}

The key parameters are of course the beam currents, beam-beam tune
shifts and the $\beta_y^*$. To reach a luminosity of 8
$\times$10$^{35}$cm$^{-2}$s$^{-1}$ (a factor of 47 higher than the
achieved one) stored currents need to increase from the values
achieved in LER and HER (1.7 A x 1.4 A) to 9.4 A x 4.1 A (a factor
5.5 and 2.9 respectively). The beam-beam parameter should go from
the achieved 0.059 to 0.24 (a factor of 4 increase), while the
$\beta_y^*$ needs to be squeezed down from 6.5 mm and 5.9 mm to 3
mm, with a simultaneous shortening of the bunch length to 3 mm.
According to beam-beam simulations this can be done if a specific
luminosity per number of bunches larger than
22$\times$10$^{30}$cm$^{-2}$s$^{-1}$ A$^{-2}$ with the crab cavities
is achieved (a factor of 2 larger than the present one at least),
and high specific luminosity at high currents(9.4 A at LER) can be
kept. Moreover 5000 bunches need to be stored, no ECI
 should arise and the bunch-by-bunch feedback system
should work without any problem.

To get the Super-KEKB design parameters the ARES copper cavities
need to be upgraded with higher energy storage ratio to support higher
currents. Superconducting cavities need to be upgraded too, in order to absorb more
HOM power up to 50 kW. The beam pipes and all vacuum components will
be replaced with higher-current-proof design. Compatibility with
SuperB design has also been explored: the arc cell lattice of the
KEKB LER can be modified to decrease the emittance to 12 nm by
weakening the magnetic field of the dipoles. Lower emittance can be
reached if the dipoles are replaced. There is no need for changing
other components, like beam pipes or geometry, but of course the IR
must be rebuilt. The KEKB HER emittance is not reduced, but unequal
emittance may be fine for operation.

\subsection{Crab Cavity Operation at KEKB}

\begin{figure}[h]
\centering
\includegraphics[width=70mm]{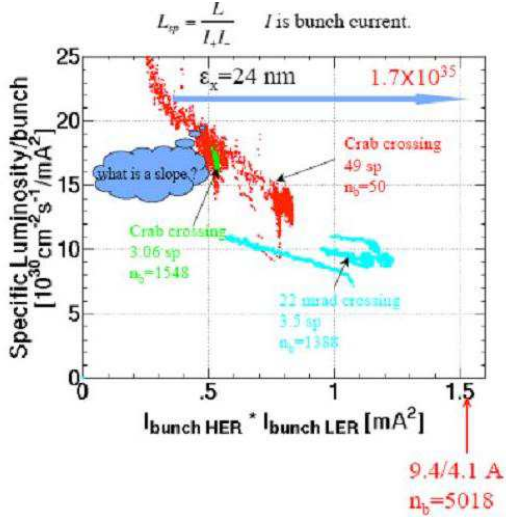}
\caption{Specific luminosity vs product of currents for different
operation scenarios.} \label{Fig3}
\end{figure}

Two crab cavities, one per ring, have been installed in KEKB last
year. The expected increment in peak luminosity, given by the
strong-strong beam-beam simulations, was about a factor of 2.
However, as it can be seen in Fig. \ref{Fig3}, where the specific
luminosity is plotted as a function of the product of beam currents
with and without crab cavities, a very high specific luminosity is
reached at low currents, dropping down faster than without crab
cavities for high beam currents. Studies are in progress to understand the causes of this
behaviour which prevents increasing luminosity at high currents.
\\

\subsection{Super-KEKB Summary}
A high current scheme approach will allow to get a luminosity for
KEKB upgrade from 5 to 8 $\times$10$^{35}$cm$^{-2}$s$^{-1}$. In case
needed, a smaller emittance of 12 nm in LER can be feasible without
hardware changes, and about 2 nm are achievable if the bends are
replaced. The design of the new vacuum system, needed to deal with
the very high stored currents, is almost completed except for the IR
chamber. In the IR design there are still things to be fixed,
especially the cure of synchrotron radiation fans on the beam pipes.
The injection complex needs also to be upgraded.

\section{SuperB}

SuperB aims at the construction of an asymmetric e$^{+}$e$^{-}$
Flavour Factory with very high peak (10$^{36}$cm$^{-2}$s$^{-1}$)
and integrated luminosity (75 ab$^{-1}$ in 5 years), with
possible location at the campus of the University of Rome Tor
Vergata, near the INFN Frascati National Laboratory (Italy). Since
2005 several Workshops have been held to prepare the Physics case,
the BaBar detector upgrade and the design of the accelerator. Many
schemes have been studied, from an ILC-like layout to a SLC-like
one. Finally, an innovative idea for collisions, supported by beam-beam simulations,
has shown the possibility to have the usual two rings scheme.
The new design is based on the ``large Piwinski angle and crab waist"
collision scheme which will allow to reach unprecedented luminosity
with low beam currents and reduced background at affordable
operating costs. The so called ``crab waist" transformation,
by means of a couple of sextupole magnets for each ring, will add a bonus
for the suppression of
synchro-betatron resonances arising from the large collision angle.
A polarized electron beam will allow for producing
polarized $\tau$ leptons, opening an entirely new realm of
exploration in lepton flavor physics. The principle of operation of
this scheme is presently under test at the DA$\Phi$NE Frascati
$\Phi$-Factory.

In its final layout the accelerator consists of two rings of
different energy (4 x 7 GeV) colliding in one IR at a large
horizontal angle. Spin rotator sections in the HER will provide
helicity of a polarized electron beam. A Conceptual Design Report
(CDR) \cite{CDR}, was issued in May 2007, with about 200 pages
dedicated to the accelerator design.

\subsection{A New Idea for Luminosity Increase}

The key point of the SuperB design is to focus more the beams at IP
and have a large crossing angle: this translates into having a large
Piwinski angle.

In summary, the design is based on:

\begin{itemize}
      \item large Piwinski angle;
      \item ``crab waist" scheme (with no RF cavities but sextupoles);
      \item very small $\beta_y^*$ at IP;
      \item small collision area;
      \item small power consumption.
\end{itemize}

\begin{figure}[h]
\centering
\includegraphics[width=60mm]{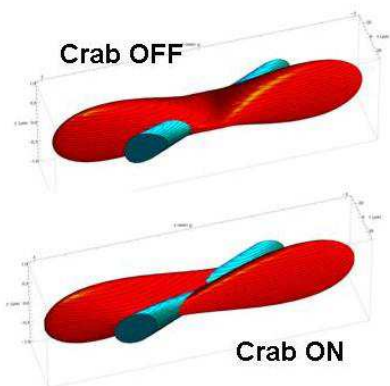}
\caption{Beam cross sections at the IP without (top) and with
(bottom) ``crab waist" transformation.} \label{Fig4}
\end{figure}

Due to the smaller collision area, it is
possible to get lower $\beta_y^*$  values without shortening the bunch length.
Moreover, due to the large crossing
angle, there will be fewer or no parasitic crossings. Two sextupoles per ring, in
phase with the IP in x and at 90 degrees in y, will suppress the
dangerous betatron and synchro-betatron resonances, and all
particles in each beam will collide at the minimum $\beta_y^*$
region (waist) with a net luminosity gain (see Fig. \ref{Fig4} for
the beam distributions without (top) and with (bottom) ``crab waist"
transformation).
As a result a higher luminosity will be possible, with same currents
and bunch length as in the present B-Factories, this means that:

\begin{itemize}
      \item beam instabilities are less severe;
      \item HOM heating is manageable;
      \item there will be no coherent synchrotron radiation (CSR) from short
      bunches;
      \item stored currents small will be smaller (less than 2 A per beam);
      \item power consumption will be much lower;
      \item background rates will be lower.
\end{itemize}

\begin{figure}[h]
\centering
\includegraphics[width=60mm]{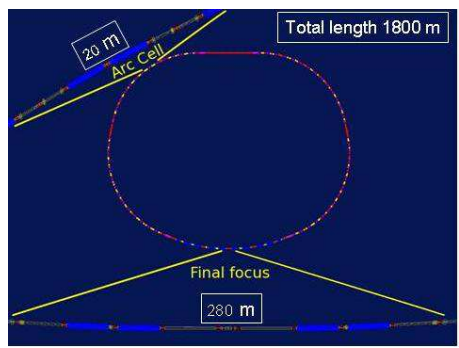}
\caption{Layout of one SuperB ring.} \label{Fig5}
\end{figure}

\begin{figure}[h]
\centering
\includegraphics[width=60mm]{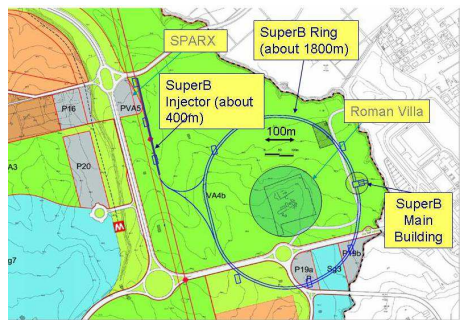}
\caption{Sketch of the SuperB accelerator on the Tor Vergata
campus.} \label{Fig6}
\end{figure}

The SuperB, as described in the Conceptual Design Report, is the
result of an international collaboration between experts from BINP,
Cockcroft Institute, INFN, KEKB, LAL/Orsay, SLAC. The design is
flexible but challenging and the synergy with the ILC Damping Rings,
which helped in focusing key issues, will be important for
addressing some of the topics (low emittance tuning, ECI remediation, etc...). No wigglers
will be needed to reach the design emittances and damping times. The
design is based on recycling all PEP-II hardware: dipoles,
quadrupoles, sextupoles, RF system, and possibly vacuum system,
allowing to reduce costs. Background studies have been carried out
in synergy with the detector experts, in order to optimize the
collimators set for backgrounds reduction. The design of the Final
Focus has been optimized in terms of chromatic corrections and
luminosity performances. The large crossing angle geometry allows
for having two separate QD0 for HER and LER: since the mechanical
constraints are too tight for a conventional septum magnet, a novel
concept to compensate the cross-talk among the two QD0's core and
fringe fields has been studied \cite{EP}. Longitudinal polarization
for the electron beam will also be included, with the possibility to run
at lower energy ($\tau$) with a loss of a factor of 10 in
luminosity. The layout of one ring is shown in Fig. \ref{Fig5},
while the possible location on the Tor Vergata University campus is
shown in Fig. \ref{Fig6}. In Table \ref{T2} is a comparison of
SuperB and Super-KEKB main parameters.

\begin{table*}[t]
\begin{center}
\caption{Comparison of SuperB to Super-KEKB.}
\begin{tabular}{|l|c|c|c|}
\hline \textbf{Parameter} & \textbf{ Units} & \textbf{ SuperB} &\textbf{ Super-KEKB}  \\
\hline Energy & GeV & 4x7 & 3.5x8 \\
\hline Luminosity & $\times$10$^{36}$cm$^{-2}$s$^{-1}$
 & 1 to 2 & 0.5 to 0.8 \\
\hline Beam currents & A & 1.9x1.9 & 9.4x4.1 \\
\hline $\beta_y^*$ & mm & 0.22 & 3. \\
\hline $\beta_x^*$ & cm & 3.5x2. & 20. \\
\hline Crossing angle (full) & mrad & 48.  & 30. to 0. \\
\hline RF power (AC line)& MW& 17. to 25.  & 80. to 90.\\
\hline Tune shifts & (x/y)& 0.0004/0.2& 0.27/0.3\\
\hline
\end{tabular}
\label{T2}
\end{center}
\end{table*}

\subsection{SuperB Summary}

SuperB has very ambitious goals in terms of peak and integrated
luminosity, supported by a new collision scheme and confirmed by
beam-beam simulations. The initial design meets the goals requested
by the experimenters. The test on this scheme is in progress at
DA$\Phi$NE and encouraging results have been achieved at the moment. The work on
the accelerator is continuing to focus on possible issues. The next
step will be to form a team to complete a Technical Design Report by
2010.

\section{Conclusions}

Operation of present B-Factories has been very successful and an
upgrade of is desirable and feasible. KEKB and PEP-II experience was
highly positive and instructive, but going to higher luminosities is
much more challenging: the ``brute force" approach seems hard to
pursue and new ideas need to be tested. Solutions to problems can
come from the collaboration between international laboratories, as
it is done for the ILC.

Two different approaches are being considered for Super-KEKB and
SuperB, with different challenges. Super-KEKB is the natural
continuation of KEKB, studies are advanced and it is waiting for
funding. SuperB exploits new concepts in colliding beams physics,
allowing for the collection of a larger data sample. The test of the
novel collision scheme is in progress and the first results of the
upgraded DA$\Phi$NE are very encouraging and important for the very
high luminosity regime required by future Flavour Physics studies.

\end{document}